# Safety Message Power Transmission Control for Vehicular Ad hoc Networks

[1]Ghassan Samara, [1]Sureswaran Ramadas and [2]Wafaa A.H. Al-Salihy
[1]National Advanced IPv6 Center,
[2]School of Computer Science,
University Sains Malaysia Penang, Malaysia

**Abstract: Problem statement:** Vehicular Ad hoc Networks (VANET) is one of the most challenging research area in the field of Mobile Ad Hoc Networks. **Approach:** In this research we proposed a dynamic power adjustment protocol that will be used for sending the periodical safety message. (Beacon) based on the analysis of the channel status depending on the channel congestion and the power used for transmission. **Results:** The Beacon Power Control (BPC) protocol first sensed and examined the percentage of the channel congestion, the result obtained was used to adjust the transmission power for the safety message to reach the optimal power. **Conclusion/Recommendations:** This will lead to decrease the congestion in the channel and achieve good channel performance and beacon dissemination.

## INTRODUCTION

VANET has attracted a wide range of research effort these days, aiming to reach road safety, infotainment and a comfort driving experience, all these benefits in low cost.

In VANET all vehicles share and compete for one 10 MHz control channel (5.885-5.895 GHz, channel 178) (Miček and Kapitulik, 2009), this channel is used for safety related messages and service announcements, each vehicle send beacons 10 times per 1 sec which will cause a heavy load on the channel. Therefore, all vehicles will have to monitor the control channel often enough to receive all safety related information so that the safety applications achieve their goal.

Safety message needs to be transmitted all the time for all near neighbors, to give information about the current status of vehicle and to let other vehicle aware about the status of near network, this critical information must be sent with high probability and reliability to avoid network problems.

In order to send the safety message in high reliability and availability some conditions must be checked before transmission to make sure that this message will reach its destination and it will not cause channel congestion, these conditions like transmission power, message size, network status and message repetition.

Sending safety message without using a congestion control mechanism creates the broadcast storm problem.

In some cases message loss rates caused by MAC collision is between 20 and 40% (Mak *et al.*, 2005).

The power limits prescribed by the Federal Communications Commission (FCC) for DSRC spectrum are as high as 33 dBm (Guan *et al.*, 2007) for vehicle on board units, so that a desired communication range of 300 m for these safety messages can be easily reached in one hop. We must take into consideration that sending safety message in maximum power, will not guarantee that the message will reach for all the vehicles on road, but guarantee to cause congestion. Trying to reach a fixed transmission power for VANET is not practical due to high mobility and large variation of distances between vehicles.

In this study we concerned with design a new protocol that will enable each vehicle on the road to automatically adjust the transmission power, which will help the network to avoid congestion caused from periodic safety message, we also analyzed the current research efforts in area of power control of safety message transmission of VANET and we are addressing our proposed protocol that contains solutions for current system.

**Analysis of relevant research area:** Many papers introduced the idea of how to reduce the channel congestion in many ways.

Mittag *et al.* (2009) authors presented a framework for a fair comparison between single hop transmission at high transmit power and multi-hop transmission and

**Corresponding Author:** Ghassan Samara, National Advanced IPv6 Center, University Sains Malaysia Penang, Malaysia

relaying at lower transmit power to know whether an efficient multi-hop beaconing can reduce the load on the channel and found that single hop must be used for beaconing and multi hop could be used for full coverage, as mentioned earlier broadcasting in full power will produce a broadcast storm problem.

Chigan and Li (2007) proposed a Delay-Bounded Dynamic Interactive Power Control (DB-DIPC), in which the transmission powers of VANET nodes are verified iteratively and interactively by the neighbor vehicles at run-time. The resulting dynamic transmission power adjustment for communications between immediate neighbor vehicles ensures that the 1-hop neighbor connectivity at run-time to adapt the high VANET dynamics promptly.

Guan *et al*. (2007) authors developed a power control algorithm to determine the transmission power for reliable vehicle safety communication by adding a power tuning feedback beacon during each safety message exchange. They found that the more data traffic loads on the channel, the greater the potential for improvement to their design.

Torrent-Moreno *et al*. (2005) authors proposed FPAV, a centralized power control algorithm that provides a solution to adjust the channel load in VANET environments problem by maximizes the minimum transmission range for all nodes in a synchronized approach, by analyzing the piggybacked beacon information received from neighbors.

Mittag *et al*. (2008) authors analyzed distributed strategies that control the vehicles' communication behavior in a cooperative manner to keep the beaconing load below a preconfigured threshold, the result showed that the overhead of the existing DFPAV approach can be reduced but still scales linearly with the number of nodes within carrier sense range.

## MATERIALS AND METHODS

**Proposed network:**
**Basic idea:** Each vehicle transmits a status message called beacon every 10 ms (White Paper, 2005), this beacon contains ID, position, direction, speed, time stamp, beacon interval (Abuelela and Olariu, 2009), the importance of the beacon is to give each vehicle information about current network status and to avoid traffic problems, each vehicle equipped with A GPS device to retain the current position.

**Preparing to send:** Each beacon received must be processed in order to get information about neighbor vehicles and about current network, the proposed beacon must hold information about transmission power to help the receiver to determine the suitable power for transmission Fig. 1.

| Seq | Int | TS | ELP | Pos | Spd | Dir | MaxP | MinP | PowU |
|-----|-----|----|----|----|----|----|------|------|------|

Piggybacked information
Seq: Beacon Sequence Number, Int: Beacon interval
TS: Time Stamp, ELP: Electronic License Plate
(Raya and Hubaux, 2005)
Pos: Position, Spd: Speed, Dir: Direction
MaxP: Maximum power received by vehicle
MinP: Minimum power received by vehicle
PowU: Power used by sender.

Fig. 1: Proposed beacon

The power information added is piggybacked to the current beacon used in VANET. Each message has a unique sequence number that it takes from MAC layer, according to IEEE Std 802.11 (2007) standards, a two-byte sequence control field is contained in an 802.11 MAC header and it could be used to detect collision and traffic load in the network Fig. 3 and Table 1.

Each receiver vehicle must hold and keep the sequence of received beacon in Sequence List (SL), to help it to determine the status of the networks traffic, Fig. 4.

The information received from beacons can be utilized in order to compute current network congestion, as beacon arrives if the network is not congested and will fail to reach its destination if there is something preventing it. We can compute the percentage of congestion by know how many beacons failed to be received in every second, as each vehicle must receive 10 beacons from each neighbor every 1 sec:

$$p = \frac{b}{100} \times 100\% \qquad (1)$$

So vehicle X in the previous example analyzes the received beacons, for the reception form vehicle A the percentage was 80 and 20% was percentage of failed beacons, as beacon 19 and 22 are messing (Fig. 2). Vehicle X also has to consider the distance between the two vehicle as the percentage of received beacons will decrease when the distance increases, the distance can be obtained from the deference of current position taken from GPS and the position of the sender, to make it easier to analyze this, we propose Distance Table (DT) which includes the vehicle IDs, percentage of reception and distance between sender and receiver, Table 2:

$$f = \frac{100 - p}{D} \qquad (2)$$

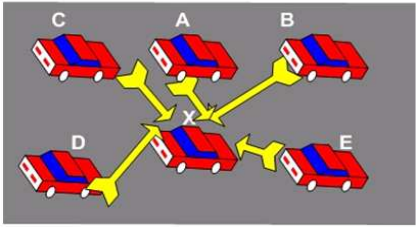

Fig. 2: Vehicle X receives beacons from neighbors

| Bytes: 2 | 4 | 6 | 6 | 6 | 2 | 2 | 0-2304+ | 4 |
|---|---|---|---|---|---|---|---|---|
| Frame Control | Duration | Address 1 | Address 2 | Address 3 | Sequence Control | QoS Control | Frame body | FCS |

Fig. 3: 802.11 MAC header (Hartenstein and Laberteaux, 2010)

Table 1: Power control algorithm parameters
| | |
|---|---|
| Receive percentage of beacons | p |
| Number of beacons received during 1 sec | b |
| Fault computed for single vehicle | f |
| Overall fault of the beacon received | F |
| Number of nodes | n |
| Percentage of receive | p |
| Distance between sender and receiver | d |
| Percentage of success for the current network status | S |
| Maximum distance for sending vehicle | MaxD |
| Minimum distance for sending vehicle | MinD |
| Power deference between max and min power | PD |
| Maximum power received from neighbors | Max BP |
| Minimum power received from neighbors | Min BP |
| Maximum power received in the field MaxP from neighbors | Ma MP |
| Minimum power received in the field MaxP from neighbors | Mi MP |

Table 2: Distance table for vehicle X
| ID | Per. of Rec. | Distance (m) | Fail |
|---|---|---|---|
| Vehicle A | 80 | 13 | 1.538 |
| Vehicle B | 60 | 18 | 2.220 |
| Vehicle C | 40 | 23 | 2.600 |
| Vehicle D | 80 | 18 | 1.110 |
| Vehicle E | 60 | 15 | 2.667 |

$$P = p - f \tag{3}$$

$$F = \sum_{n=1}^{n} \left( \frac{100 - P}{D} \right) \div n \tag{4}$$

$$S = 100\% - \left( \frac{MaxD - MinD}{2} \times F \right)\% \tag{5}$$

Returning to our example vehicle A received 8 beacons in 1 sec, from Eq. 1 p = 80%, from Eq. 2 f = 1.538, which means that 1.538 beacons fail every 1 m, so if the distance for this vehicle increases for 1 m another 1.538 beacons will be lost and the percentage of received beacon will be 78.46%.

| | | | | | | | | |
|---|---|---|---|---|---|---|---|---|
| Vehicle A | 15 | 16 | 17 | 18 | | 20 | 21 | | 23 | 24 |
| Vehicle B | 71 | 72 | | | 75 | | | 78 | 79 | 80 |
| Vehicle C | 89 | 90 | | | | | | 96 | 97 | |
| Vehicle D | 22 | 23 | 24 | 25 | 26 | 27 | | 29 | 30 | |
| Vehicle E | 61 | 62 | 63 | | | | 67 | | 69 | 70 |

Fig. 4: Sequence number received from neighbors (SL) for Vehicle X (Balon and Guo, 2006)

| Seq | Int | TS | ELP | Pos | Spd | Dir | Maxp | MinP | PowU |
|---|---|---|---|---|---|---|---|---|---|
| 15 | 50 | 1.02 | A | N05 E100 | 60 | E | 28 | 24 | 25 |
| 71 | 50 | 1.02 | B | N08 E105 | 80 | E | 29 | 23 | 28 |
| 89 | 50 | 1.02 | C | N10 E115 | 70 | E | 28 | 24 | 29 |
| 22 | 50 | 1.02 | D | N09 E106 | 50 | E | 27 | 24 | 28 |
| 61 | 50 | 1.02 | E | N06 E101 | 70 | E | 26 | 23 | 28 |

Fig. 5: Active Beacon List (ABL)

From Eq. 4 we can estimate the overall fault for the current system and it is for our example 2.027% fault for each meter and from the fifth equation we conclude that the mean percentage of successful received beacon is 63.51%.

The received beacon also includes information about power like maximum and minimum power received and transmission power used; this information is filled in Active Beacon List (ABL), Fig. 5.

From ABL vehicle X can analyze at any moment the transmission power for received beacons from neighbors, the received power depends on distance between the two parties and on the channel status, for instance, if vehicle C transmit in power less than 29, the beacon may not arrive and higher power covers wider distances and may cause much more congestion, see Fig. 6.

**Sending beacon:** Each vehicle collects its information like Speed, Direction, Position (GPS), Max power for transmission received and Min power for transmission received and power used and adds them altogether into the beacon:

$$PD = MaxBP - MinBP \tag{6}$$

$$PowU = MinBP + (PD \times S) \tag{7}$$

So from Eq. 6 the vehicle can compute the difference between the maximum and minimum power received, the importance of the two received numbers is that the minimum power received is the minimum power could be used to send and this number can be used successfully but it is may be not enough for the beacon to reach to all near neighbors and the maximum power received for the beacon as this power that help to make the congestion previously computed, so the maximum power received must be decreased in order to reduce the channel congestion and the minimum power must be increased to ensure that this beacon will arrive to further neighbors, but this increase must not exceed the maximum power received and the decrease and the increase must depend on the congestion obtained from Eq. 5. and 6 the vehicle will reach the optimal power that it should transmit its beacons using it.

For our example PD = 29-25 = 4 dBm, the network at these values suffers from congestion and these value must be changed to decrease and avoid such congestion we have to decrease the maximum power:

$$25 + 2.5404 = 27.54 \text{ dBm}$$

27.54 dBm is the optimal transmission power for this case.

```
// receive Piggyback Beacon algorithm
1.   Receive Beacon
     //Timer works every 1 second
     Timer
     {
2.   Clear DT
3.   p = b/10 * 100% // compute p
4.   f = (100-p)/D // compute f
5.   P = p - f // compute P
6.   F = (∑_{n=1}^{n} (100-P)/D) ÷ n // compute F
7.   S = (100 - ((MaxD-MinD)/2 * F)) // compute S
8.   MaxP = max power
9.   MinP = min power
10.  }
```

Fig. 6: Receive piggyback beacon

```
1.   // Sending Piggyback Beacon algorithm
2.   If S<50%
3.   {
4.   PD = MaxBP - MinBP // compute power
     deference
5.   PowU = MinP + (PD × S)   // compute
     transmission power
6.   If PowU< MaMP and PowU>MiMP
7.   {
8.   MaxP= MaxBP
9.   MinP= MinBP
10.  PowU=PowU
11.  Send
12.  }
13.  }
14.  Else
15.  Die
```

Fig. 7: Sending piggyback beacon

So for next transmission the power 27.5 dBm will be used for transmitting the beacons and this number will be updated after 1 sec when new analysis is computed for the channel status (Fig. 7).

**No congestion case:** In case that S = 100% which means that the percentage of congestion is null, this means that the maximum power received from vehicle didn't cause congestion to the channel, at this case the vehicle will compute the distance between the receiver and the vehicle that sent the higher power, if the distance is greater than 200 m, this means that the vehicle can send in maximum power received and this power will not make congestion, in another case if the distance between the receiver and the sender of the maximum power is between 100 and 200 m, this means that there may be vehicles located in the distance greater than 200 m and they are using power less than required to reach current vehicle, so the power will be used for the transmission will be:

$$PowU = MaxBP + PD \times 0.5 \qquad (8)$$

For our example Max BP was 29, from Eq. 8 PowU = 31, 31 < = 33 d B m so this power is acceptable and can reach more than 200 m. in the third case where the distance is less than 100 m, the power will be:

$$PowU = MaxBP + PD \qquad (9)$$

For our example Max BP was 29, from Eq. 9 PowU = 33, 33 <= 33 dBm so this power is acceptable and can reach more than 100 m (Fig. 8), that contains pseudo code about no congestion case.

```
// no congestion case algorithm
1.   If s = 100%
2.   {
3.   If d > 200
4.      PowU = MaxBP
5.   Else if d > 100 and (MaxBP + PD * 0.5 <= 33)
6.      PowU = MaxBP + PD * 0.5
7.   Else
8.      PowU= 33
9.   Else if d <= 100 and ( MaxBP + PD <= 33)
10.     PowU = MaxBP + PD
11.  Else
12.     PowU= 33
13.  }
```

Fig. 8: No congestion case

## RESULTS AND DISCUSSION

Maximum power allocated from ITS (33 dBm) could be used in safety message transmission and this power theoretically enables the safety message to reach 300m in best conditions, but best conditions rarely happen and congestion happen in most of the time and trying to send any message in high power in presence of the congestion will make the situation worse and the problem bigger, starting from this point the channel congestion must be detected in order to use the suitable power to ensure that the message will reach its destination and channel congestion will be reduced.

Channel congestion is computed in Eq. 1-5. These equations analyze the channel congestion in every 1 second, as each vehicle must receive 10 beacons from each neighbor; channel congestion status then is utilized to adjust the transmission power, if there is congestion, this means that maximum power used in the channel will increase this congestion, so this power must be decreased in order to reduce the overhead caused by maximum power used for transmission in the channel, this can be done using Eq. 6 and 7, the benefit of minimum power used that this is the minimum power could be used in order to ensure that the message will reach to its distention, lower power will not guarantee the reachability of these messages, this dynamic adjustment for transmission power, guarantees to reach to the optimal power that must be used.

## CONCLUSION

Safety message providing critical and important information for every vehicle on the road that must be sent all the time to make all the vehicles aware about the status of their neighbors, but sending this message causes network overhead and channel congestion that must be reduced and eliminated. Reaching the suitable transmission power is important and critical in VANET, the dynamic BPC protocol for power control that decreases the channel congestion and improves the system performance depending on the channel status and on power received, in our future work we will perform the simulation for BPC protocol and compare it with other power protocols like FPAV.